\documentclass[epj]{svjour}

\usepackage{graphicx,amsmath}
\usepackage{units}
\def\be{\begin{equation}}                                                    
\def\ee{\end{equation}}                                                    
\def\pt{p_\perp}
\def\et{E_\perp}
\def\d{\text{d}}

\begin{document}
\title{Diffraction and correlations at the LHC: definitions and observables}
\author{V.A.~Khoze\inst{1,2} \and F.~Krauss\inst{1} \and A.D.~Martin\inst{1}
\and M.G.~Ryskin\inst{1,2} \and K.C.~Zapp\inst{1}}
%
%
\institute{Institute for Particle Physics Phenomenology, 
University of Durham, Durham, DH1 3LE \and 
Petersburg Nuclear Physics Institute, Gatchina, St.~Petersburg, 188300, Russia}
\date{Received: date / Revised version: date}
%
\abstract{
We note that the definition of diffractive events is a matter of convention.  
We discuss two possible ``definitions'': one based on unitarity and the other 
on Large Rapidity Gaps (LRG) or Pomeron exchange. LRG can also arise from 
fluctuations and we quantify this effect and some of the related uncertainties. 
We find care must be taken in extracting the Pomeron contribution from LRG
events. 
We show that long-range correlations in multiplicities can arise from the same
multi-Pomeron
diagrams that are responsible for LRG events, and 
explain how early LHC data can illuminate our understanding of `soft'
interactions. 
\PACS{
      {PACS-key}{discribing text of that key}   \and
      {PACS-key}{discribing text of that key}
     } 
} 
\maketitle

\section {Introduction}

For many studies at hadron colliders it is important to unambiguously define
what component of the inelastic cross section is selected. Traditionally
observables such as the single-particle inclusive cross section and multiplicity
distributions are given for non-diffractive events. These already serve as
important input to understanding high-energy strong interactions and the tuning
of Monte Carlos. On the other hand, it is not so clear to what extent
non-diffractive processes can be disentangled. Recall, first, that inelastic 
diffraction is responsible for a sizeable part (say, about $0.2-0.3$) of the 
total $pp$ cross section; second, that the present LHC detectors do not have 
$4\pi$ geometry and do not cover the whole available rapidity interval.
So the minimum-bias events account only for a {\it part} of the total inelastic 
cross section \cite{mb}.  The extrapolation necessary to obtain the value of
the 
total cross section is {\it model dependent}, and the uncertainties associated 
with this extrapolation will limit the accuracy of the total inelastic cross 
section measurements at the LHC.  

Moreover, even when reconstructing the total cross section using the dedicated 
Totem detector \cite{totem} there is still an uncertainty due to incomplete
knowledge 
of the low-mass single diffraction (SD) contribution. This, in turn, imposes 
restrictions on the accuracy of luminosity determination using the optical
theorem, 
see for instance \cite {lumi}.

Note also that knowledge of the total inelastic cross section is important for 
the evaluation of quantities such as the number of interactions per bunch
crossing, 
when a high luminosity of the LHC becomes available.

On the theory side, the situation is not so clear.  At the moment the
theoretical 
predictions for the total $pp$ cross section, $\sigma_{\rm tot}$, at the LHC
energy 
of 14 TeV differ by a factor of 2.5 in the range between 90 and 230 GeV; 
recent 
reviews and references can be found in \cite{block,Fiore}.  Even models that
are 
ideologically close \cite{nnsoft,GLMM,kaid,ostap}, which incorporate (both
eikonal 
and enhanced) absorptive corrections, differ in their predictions for 
$\sigma_{\rm tot}$ at the LHC by about 30\% (covering the range of
90-130 mb),
while the expectations for $\sigma_{\rm SD}$ in these models differ by
a factor of 1.7 (in the range 11-19 mb).

\section{Definition of diffraction}

At the outset, we have to say there is no unique definition of diffraction.  It
is 
a matter of {\it convention} only. Usually when talking about diffraction we 
mean events arising from Pomeron exchange, which experimentally we would like
to 
associate with Large Rapidity Gaps (LRG). Unfortunately the situation is more 
complicated.  LRG can also arise from the secondary Reggeon exchange or from 
simple fluctuations in the distribution of secondaries produced in the event. 
We amplify the problem by discussing two ``definitions'' of diffraction in 
detail. The second definition will be based on the association of Pomeron
exchange 
with LRG (see Section \ref{sec:second}), but, first we consider the 
unitarity-based definition which, at first sight, looks as if it may be unique.

\subsection{First ``definition'' of diffraction}

\begin{figure*}
\begin{center}
\includegraphics[height=8cm]{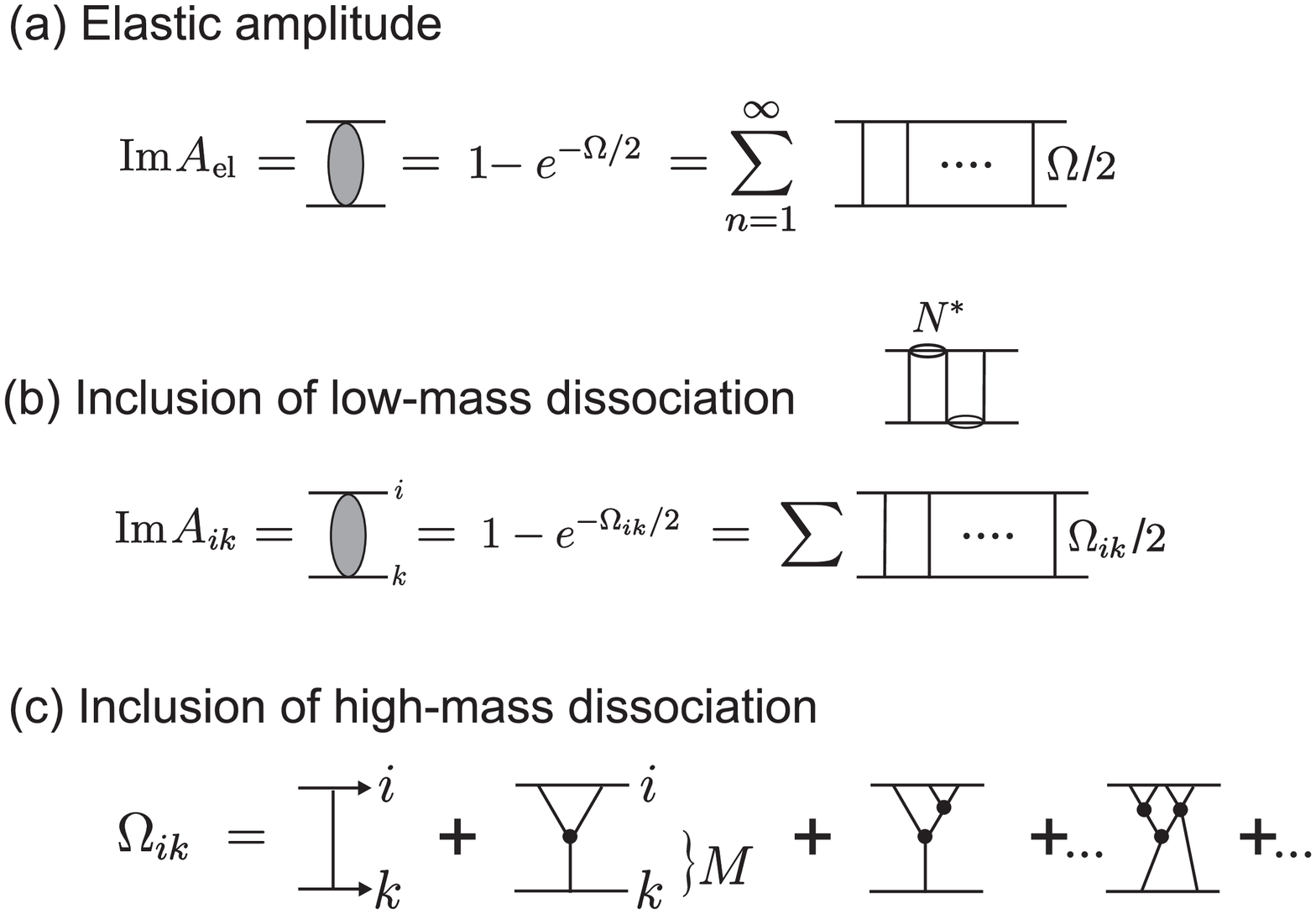}
\caption{\sf 
(a) The single-channel eikonal description of elastic scattering; 
(b) the multichannel eikonal formula which allows for low-mass proton
dissociations 
in terms of diffractive eigenstates $|\phi_i\rangle,~|\phi_k\rangle$; and 
(c) the inclusion of the multi-Pomeron-Pomeron diagrams which allow for
high-mass 
dissociation. In all these diagrams the exchanged lines represent Pomeron
exchange. }
\label{fig:epip}
\end{center}
\end{figure*}

In analogy with optics, we could say that {\it diffraction} is {\it ``elastic'' 
scattering caused, via unitarity, by the absorption of components of the wave 
functions of the incoming protons}.  If we just consider elastic unitarity,
then 
it gives an elastic amplitude of the form \cite{softLHC}
\begin{equation}
A(b)=i(1-\exp(-\Omega(b)/2))
\label{eq:eik}
\end{equation}
in impact parameter, $b$, space, where $\Omega$ is the opacity or eikonal. 
The situation is sketched symbolically in Fig. \ref{fig:epip}(a).

Note, however, that as a rule the wave functions of incoming hadrons differ
from 
the eigenstates corresponding to high energy scattering amplitude. Different 
components of the initial hadron wave function have different absorptive cross
sections. 
As a consequence, the  outgoing superposition of states will be different from 
the incident  particle, so we will have {\it inelastic}, as well as elastic, 
diffraction.

To discuss inelastic diffraction, it is convenient to follow Good and 
Walker \cite{GW}, and to introduce states $\phi_k$ which diagonalize the $A$ 
matrix. These, so-called diffractive, eigenstates only undergo `elastic' 
scattering. To account for the internal structure of the proton we, therefore, 
have to enlarge the set of intermediate states, from just the single elastic 
channel, and to introduce a multi-channel eikonal, see Fig. \ref{fig:epip}(b).

This definition of diffraction is good for elastic and quasi-elastic processes, 
where the size of the LRG is close to the whole available rapidity interval.
But 
what about proton dissociation into high-mass systems? At first sight, it
appears 
that we may account for it by simply enlarging the number of diffractive 
eigenstates $\phi_k$.  Unfortunately, this does not work. To see this, we
decompose 
the wave functions of the incoming protons in terms of parton distributions.  
At very small $x$, where we sample the `sea' parton distributions, we face a 
difficulty. The problem is double counting, which occurs when the `sea' partons 
originating from the dissociation of the `beam' and the `target' protons overlap
in 
rapidities, as shown in Fig. \ref{fig:loops}(a); see  \cite{nsoft}. 

Thus, to avoid double counting, a reasonable convention for assigning the
parton 
contributions would be to build up the Good-Walker diffractive eigenstates just 
from the valence distributions, and to attribute the sea distributions to
Pomeron 
exchange. The multi-Pomeron exchange diagrams (such as those of 
Fig. \ref{fig:epip}(c)) describe, among other things, large rapidity gap (LRG) 
events. The last component is often called high-mass diffraction\footnote{
  We emphasize that each multi-Pomeron exchange diagram describes several 
  different processes, according to whether or not the individual Pomeron 
  exchanges are `cut'. The LRG component corresponds to the case, where in 
  some rapidity interval, the cuts go between the Pomerons, so that there 
  are no particles produced in this rapidity interval.}. 
Thus, in a narrow sense, {\it diffraction} may be defined as just the elastic 
scattering of the Good-Walker eigenstates, but a more general definition would
be 
to include the LRG processes arising from the multi-Pomeron interactions.
Hence, 
we now turn to a definition based on LRG.

\subsection{Second ``definition'' of diffraction \label{sec:second}}

 Another possibility is to call {\it diffractive} any process 
{\it caused by Pomeron exchange}, or to be more precise - by the exchange 
corresponding to the `rightmost vacuum singularity' in the complex angular 
momentum plane.  It was the old convention that any event with a LRG of the 
size $\delta \eta>3$ may be called ``diffractive'', since here Pomeron exchange 
gives the major contribution. Unfortunately, again, the situation is more
complicated.  
LRG, in the distribution of secondaries, may occur due to the Reggeon exchange, 
when the colour flow across the gap is absent\footnote{
  We are not discussing here rapidity gaps caused by electroweak exchanges.} 
and the parton wave function saves its initial coherence within the interval 
occupied by the Reggeon (that is across the LRG), see Section
\ref{sec:reggeon}. 
In addition, a gap may also occur just due to fluctuations of secondaries
generated 
during the hadronization process. Indeed, we show in Section \ref{sec:fluc}
that 
events with LRG of size $\delta \eta >3$ do not unambiguously select
diffractive 
events at LHC energies.

\section{LRG from Reggeon and Pomeron exchange \label{sec:reggeon}}

\begin{figure*} [t]
\begin{center}
\includegraphics[height=5cm]{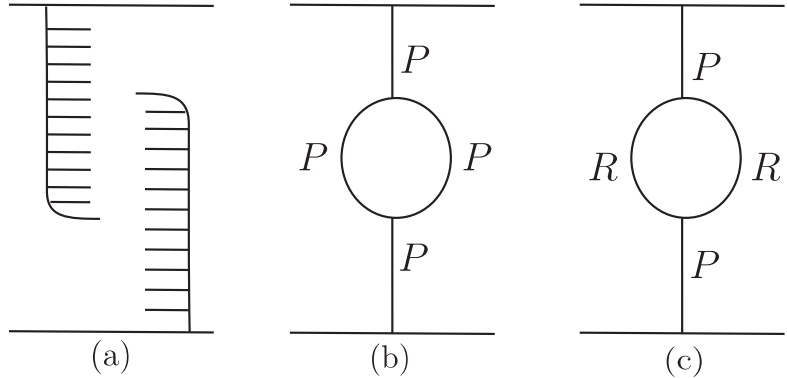}
\caption{\sf 
  (a) A `soft' high energy interaction in which the sea partons which originate 
  from the dissociation of the colliding protons overlap in rapidity. The
overlap 
  illustrates the impossibility of achieving a unique definition of diffraction
in 
  terms of Good-Walker diffractive eigenstates; 
  (b) is a corresponding Pomeron loop diagram. 
  Plot (c) is a Reggeon loop diagram. There are four different cuts of the 
  Pomerons in the loop of diagram (b): we may cut either the left or the right 
  or neither or both of the Pomerons corresponding to processes of diagram (a) 
  where the coherence of the partons in the central (`overlapping') rapidity
region 
  is destroyed or saved in the right or left parton shower. A cut of a Pomeron 
  means that the coherence of the corresponding parton shower is destroyed. A 
  rapidity gap occurs when neither Pomeron in the loop is cut.}
\label{fig:loops}
\end{center}
\end{figure*}
First, we discuss the gaps caused by Reggeon exchange.  In terms of Regge Field 
Theory (RFT) \cite{RFT} this contribution is described by the Reggeon loop, see 
Fig. \ref{fig:loops}(c).   If we neglect the multi-Reggeon diagrams, then the 
corresponding probability to get LRGs takes the form
\begin{equation}
 P_R(\delta \eta)=c_R\exp(\delta \eta(2\alpha_R-\alpha_P-1)),
\end{equation}
where $\alpha_R$ is the trajectory of the secondary Reggeon, while $\alpha_P$
is 
the intercept of the vacuum singularity (Pomeron).

On the other hand for the Pomeron loop, Fig. \ref{fig:loops}(b), the probability
is
\begin{equation}
 P_P(\delta \eta)=c_P\exp(\delta \eta(\alpha_P-1)).
\end{equation}
Thus, secondary Reggeon exchange\footnote{
  Here we put $\alpha_R=1/2$.} 
produces gaps with a correlation length $l_R = -1/(2\alpha_R-\alpha_P-1)\sim 1$,
while the Pomeron loop provides a long range correlation with length 
$l_P\sim  O(1/(\alpha_P-1))$.

In the framework of perturbative QCD, Pomeron exchange is given by the BFKL 
amplitude \cite{bfkl}. After accounting for the resummed Next-to-Leading Log
(NLL) 
corrections \cite{nll,resum,resum1} the expected BFKL Pomeron intercept is 
$\alpha_P=1+\Delta_P\sim 1.3$. On the other hand, the present data on
high-energy 
total cross sections are well described by a `soft' Donnachie-Landshoff 
parametrization \cite{dl}, corresponding to single Pomeron exchange (without
the 
multi-Pomeron contributions) with an `effective' 
$\alpha_{\rm DL}(0)-1=\Delta_{\rm DL}=0.08$.  

There are two interpretations for the reduction in the rise of the cross
section,  
$\sigma_{\rm tot}\propto s^\Delta$,  from $\Delta=0.3$ to $\Delta=0.08$:  
\begin{itemize}
\item
  If it is caused by eikonal-like screening corrections, (\ref{eq:eik}), that is
by the 
  rescatterings of the initial fast hadrons, while the intercept corresponding
to an 
  individual parton shower (the Pomeron) remains large ($\Delta\sim 0.3$), then
the 
  major part of the LRG events are those events with the {\it largest} available
  
  rapidity gaps (the probability $P_P(\delta \eta)$ grows with $\delta \eta$). 
Therefore, in
  this framework, we actually come back to low-mass, quasi-elastic, diffractive 
  dissociation. Conversely, high-mass dissociation will give a small
contribution only. 
\item
  Alternatively, if the effective Pomeron intercept becomes smaller due to 
  rescattering (screening)  between the internal partons inside an individual
parton 
  cascade (the BFKL ladder) (thought to be due to the so-called enhanced
diagrams 
  \cite{cardy}), then the probability of high-mass diffractive dissociation will
be 
  large. In addition, then the corresponding correlation length is expected to
be 
  of the order of the `renormalized' $1/\Delta\sim 10$.
\end{itemize}

Besides this, there may be an interference contribution arising from the
secondary 
Reggeon across the gap in the amplitude $A$ and the Pomeron exchange in the
amplitude 
$A^*$ (and vice versa). Then the  corresponding correlation length is 
$l_{\rm int}=1/(1-\alpha_R)\sim 2$.  Obviously, this additional effect will blur
the
two alternative pictures outlined above.

\section{LRG from fluctuations \label{sec:fluc}}
In addition, there is yet another effect, which will further obfuscate the
picture:
Obviously, a rapidity gap may also occur just due to fluctuations of
secondaries 
generated during the hadronization process in an otherwise 'perfectly
inelastic' 
event.  We will first crudely estimate the probability of LRGs arising from
fluctuations 
at the LHC, before we perform a more detailed Monte Carlo study of this
possibility.

\subsection{Analytic estimate}

If we assume an independent creation of secondaries in each rapidity interval,
then 
the probability to get a gap during hadronization can be written as
\begin{equation}
P_{\rm fluc}(\delta \eta)\ =\ \frac 1{l_f} \exp(-\delta \eta/l_f).
\label{hadr}
\end{equation}
The distribution over the gap size measured at the Tevatron by CDF \cite{cdf} 
indicates that the correlation length $l_f\sim 0.7- 0.75$.

To obtain the corresponding cross section, the probability (\ref{hadr}) should 
be multiplied by the expected inelastic cross section $\sigma_{\rm inel}\sim 50$
mb 
and integrated over the chosen interval in $\eta$ with a weight equal to the 
particle density $\d N/\d\eta$. Thus, the cross section of the events with a
gap $\delta \eta$
larger than $[\delta\eta]_{\rm min} \equiv \Delta\eta$ is
\begin{equation}
\sigma_{\rm fluc}(\delta \eta>\Delta \eta)=
\sigma_{\rm inel}\int P_{\rm fluc}(\delta \eta)~\d\delta \eta~\d\eta
\frac{\d N}{\d\eta}.
\end{equation}
For an  estimate at the LHC, we take $l_f=0.7$ and\footnote{
Here $N$ includes, not only charged, but neutral hadrons as well.} 
$\d N/\d\eta\sim 3$.
We use the ATLAS cuts \cite{ATLAS} of $\pt>0.5$ GeV and $|\eta|<5$. We find
the probability
\be
P_{\rm fluc}(\delta\eta>3)~=~\exp(-3/0.7)\cdot(\d N/\d\eta)\cdot \Delta\eta \sim
0.25,
\ee
which with $\sigma_{\rm inel}\sim 50$ mb gives  
$\sigma_{\rm fluc}(\delta \eta>3)\sim 10$ mb. This value should be compared with
the 
expected cross section of diffractive double dissociation $\sigma^{\rm DD}\sim
10$ mb, 
which according to the data from lower energy colliders, has a very flat energy 
dependence \cite{cdf}.

\subsection{Monte Carlo study of LRG from fluctuations}

In order to obtain a somewhat more realistic estimate of fluctuation
contributions to
different event classes, inclusive QCD events were generated using the
\textsc{Sherpa}~1.2.1 event generator~\cite{Gleisberg:2008ta}.  This sample
contains no diffractive events.  On the matrix element level, partonic $2 \to 2$
scattering processes are generated, which are supplemented by a Catani-Seymour
dipole shower~\cite{Schumann:2007mg}.  Multiple interactions are simulated using
a model based on
\cite{Sjostrand:1987su,Alekhin:2005dx}, which bases on independent (semi-)hard 
partonic scattering processes with decreasing $p_\perp$.  On the pertubative
level,
the only correlation between individual parton scatters emerges through the
rescaling
of the parton distribution functions after each scatter and its parton shower
has
taken place.  The parameters of the multiple interaction model where tuned to 
describe Tevatron underlying event data. \textsc{Sherpa} by default uses a
cluster
hadronisation~\cite{Winter:2003tt}, but, it can also can also hand over events
to
the Lund string fragmentation in \textsc{Pythia}~\cite{Sjostrand:2003wg}.  Both
models
were tuned to reproduce LEP data\footnote{
  Note, though, that the tune of the Lund fragmentation performed was quite
crude.}. 
The events were analysed using \textsc{Rivet}~\cite{Buckley:2008vh}. We use
both models to get an estimate of the uncertainty. Our studies are performed at
the hadron level without accounting for detector effects.

\begin{figure*}
\begin{center}
\includegraphics{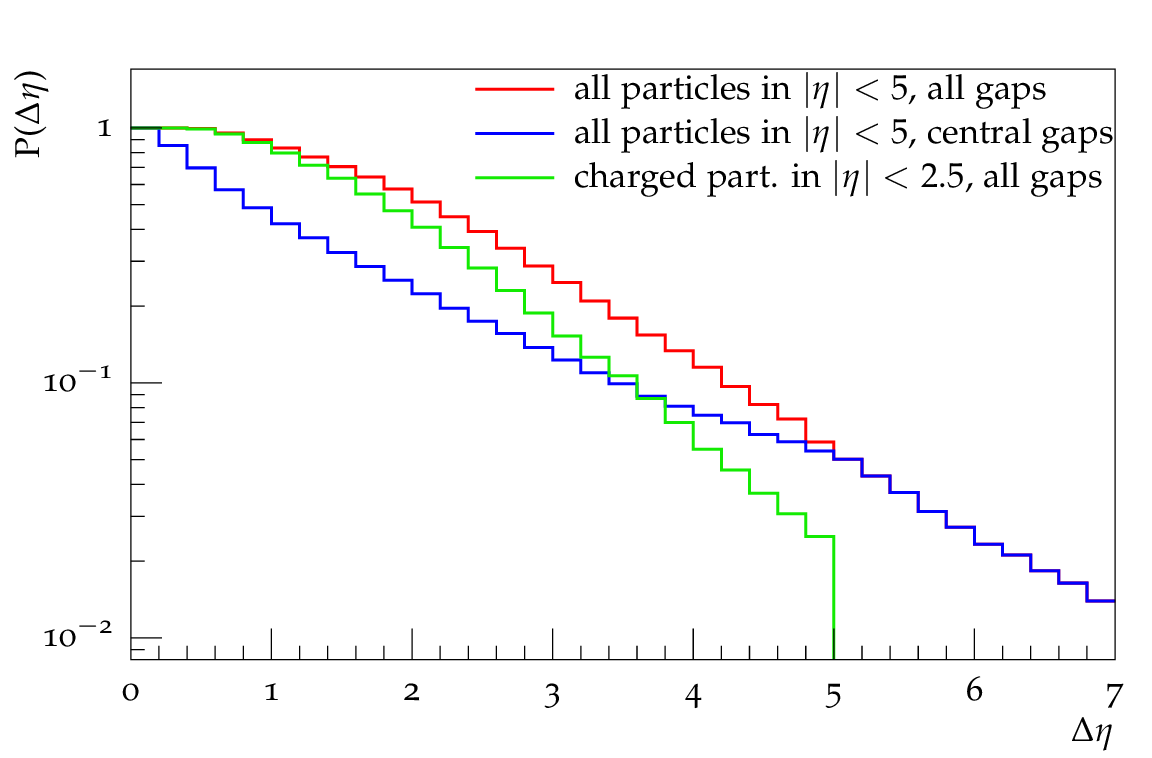}
\caption{\sf Probability for finding a rapidity gap larger than $\Delta\eta$ in
 an inclusive QCD event (cluster hadronisation) for different gap definitions.
'All gaps' refers to a scenario where the gap can be anywhere in the
acceptance, while in 'central gaps' the gap is required to be central
(i.e.\ $\eta=0$ has to lie in the gap). The $p_\perp$ threshold is
\unit[500]{MeV} and no trigger condition was required,
$\sqrt{s}=\unit[7]{TeV}$.}
\label{fig:mc_gapdef}
\end{center}
\end{figure*}

The probability for finding a gap larger than some $\Delta\eta$ in an event was
extracted as a function of the gap size for different gap definitions,
hadronisation models, threshold $p_\perp$, trigger conditions and beam
energies.  The default set-up is as follows:
\begin{itemize}
\item the hadronic c.m.-enegy is given by $\sqrt{s_{pp}}=\unit[7]{TeV}$.
\item all particles (charged and neutral) in $|\eta|<5$ with a
$p_\perp$-threshold 
  of $p_{\perp,\,\text{cut}}=\unit[0.5]{GeV}$ are considered;
\item the gap is allowed to be anywhere in the 'calorimeter acceptance' 
  $\eta\in [-5,\,5]$
\item no further trigger condition is required
\item the simulation bases on inelastic $2\to2$ scatters in QCD, with 
  $p_\perp>\unit[2.8]{GeV}$, supplemented with \textsc{Sherpa}'s default 
  parton shower and cluster hadronisation.
\end{itemize}

Fig.~\ref{fig:mc_gapdef} shows the gap probability for three different gap
definitions, which are inspired by the \textsc{Atlas} acceptance: The first
('all') is the default definition (all particles in $|\eta|<5$, the gap can be
anywhere), the second ('central') also takes all particles in $|\eta|<5$ but
requires the gap to be central (i.e.\, including $\eta=0$). In the third option
('charged') only charged particles in the tracking region $|\eta|<2.5$ are
considered and the gap, again, can be anywhere. The observed gap rates are
generally
sizable -- in case of the 'all' definition as shown in Fig.~\ref{fig:mc_gapdef}
the probability for observing a rapidity gap larger than 3 units is around
\unit[25]{\%} and the probability for a central gap with $\Delta\eta > 3$ is
still sizable, around \unit[10-15]{\%}. Given the large inelastic QCD cross 
section these are worryingly large numbers.

\begin{figure*}
\begin{center}
\includegraphics{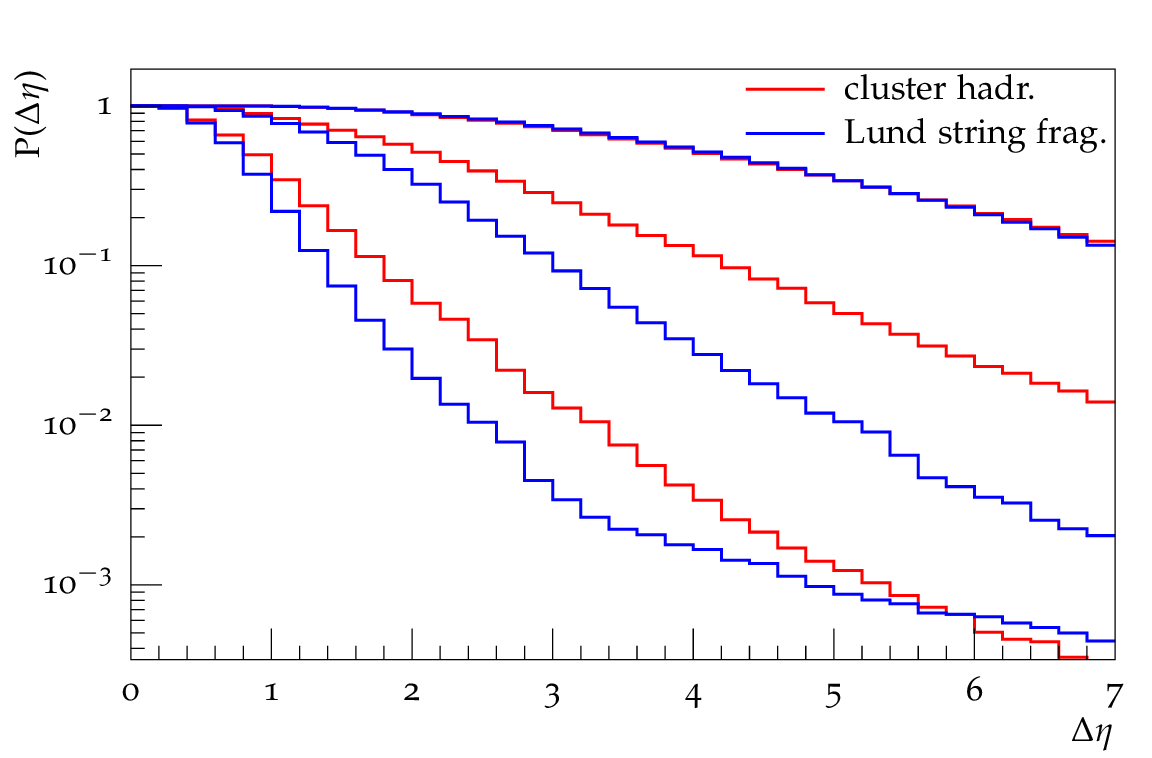}
\caption{\sf Probability for finding a rapidity gap (definition 'all') larger
than $\Delta\eta$ in an inclusive QCD event for different threshold $p_\perp$.
From top to bottom the thresholds are $p_{\perp,\,\text{cut}}=\unit[1.0\, ,\,
0.5\, ,\, 0.1]{GeV}$. Note that the lines for cluster and string hadronisation
lie on 
top of each other for $p_{\perp,\,\text{cut}}=\unit[1.0]{GeV}$. No trigger 
condition was required, $\sqrt{s}=\unit[7]{TeV}$.}
\label{fig:mc_had}
\end{center}
\end{figure*}

In Fig.~\ref{fig:mc_had} the dependence on the $\pt$ cut and the hadronisation
model is investigated. As expected, the gap probability depends strongly on
the $\pt$ cut. Since most particles are rather soft a moderate $\pt$ cut
removes a considerable fraction of the particles from the event and thus
drastically increases the chances of observing a large rapidity gap. At high
$\pt$ the cluster and string fragmentation yield identical results (upper line
in Fig.~\ref{fig:mc_had}), but at lower $\pt$ the difference between the two
models becomes sizable. Recall, however, that the string model was only crudely 
tuned together with the \textsc{Sherpa} parton shower. The discrepancies seen in
Fig.~\ref{fig:mc_had} should thus be regarded as an upper bound for the model
uncertainty due to hadronisation effects.  As an additional test, the cluster 
hadronisation model in \textsc{Sherpa} was employed with two different
algorithms of 
assigning colour in the final state,  to see if non-pertubative colour
reconnection
effects have a sizable impact.  In these two algorithms, minimising the 'length'
of
the total colour flow in momentum space and random assignment, no significant
change 
of the size of the difference between the cluster and the string fragmentation
has 
been found.  It is also noteworthy that the string hadronisation produces for 
low $\pt$ thresholds a distribution that is not exponential but has two
components 
that resemble a Pomeron contribution at large gap sizes.

\begin{figure*}
\begin{center}
\includegraphics{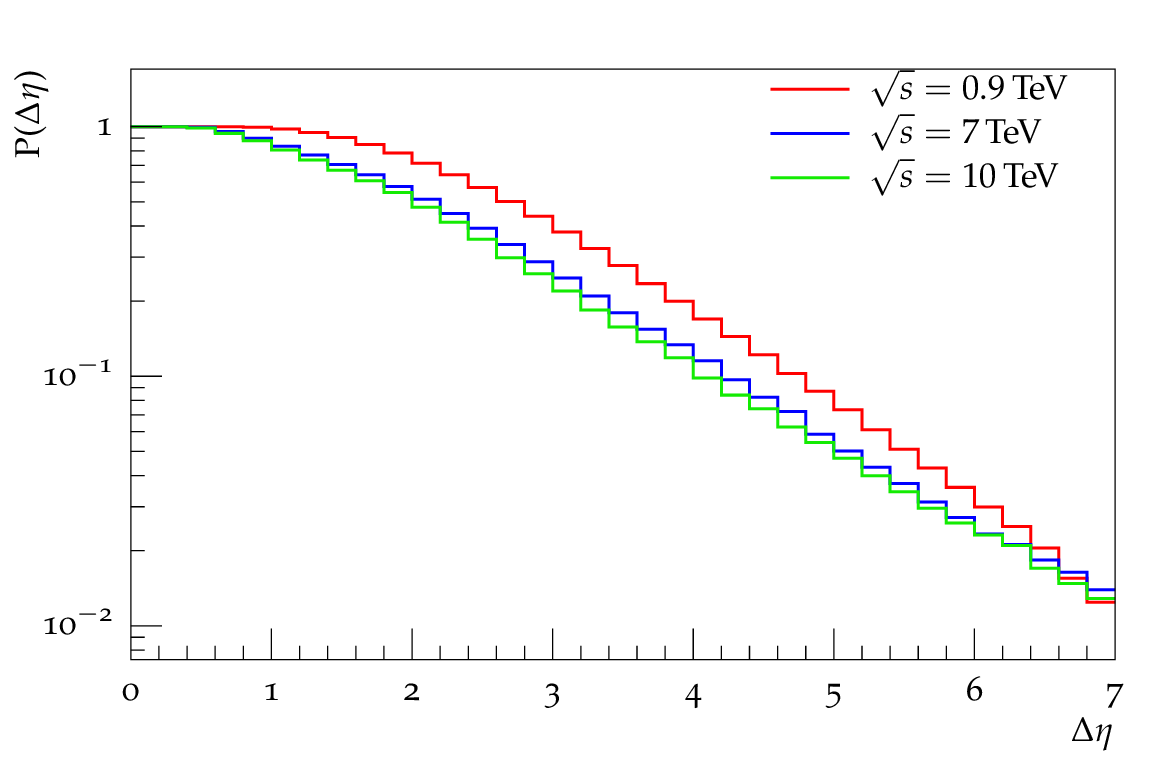}
\caption{\sf Beam energy dependence of the probability for finding a rapidity
gap (definition 'all') larger than $\Delta\eta$ in an inclusive QCD event
($p_{\perp,\,\text{cut}}=\unit[0.5]{GeV}$, no trigger condition, cluster
hadronisation). }
\label{fig:mc_energy}
\end{center}
\end{figure*}

The gap probability decreases moderately with increasing beam energy
(Fig.~\ref{fig:mc_energy}), since the multiplicity and mean $p_\perp$ increase
with $\sqrt{s}$. This translates directly into a lower probability for large
gaps caused by fluctuations.

\begin{figure*}
\begin{center}
\includegraphics{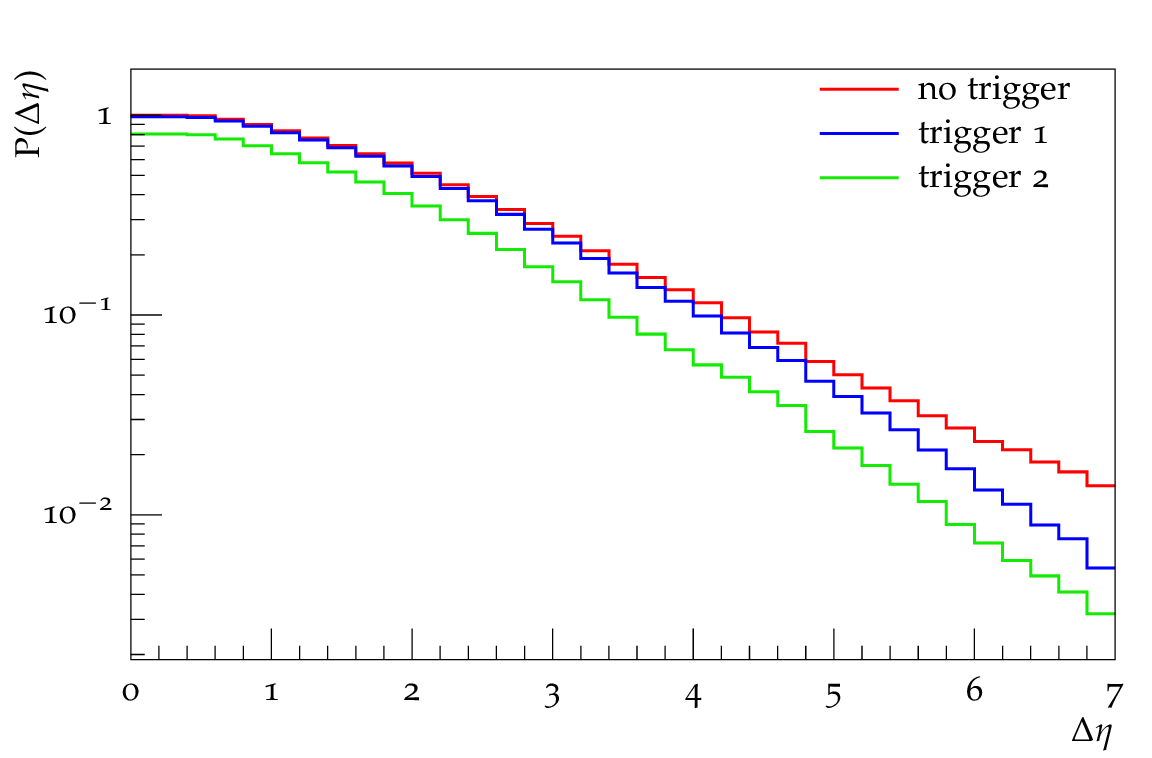}
\caption{\sf Probability for finding a rapidity gap (definition 'all') larger
than $\Delta\eta$ in an inclusive QCD event with different trigger conditions.
Trigger~1\,(2) requires at least one charged particle with $p_\perp >
\unit[0.5]{GeV}$\,($p_\perp > \unit[1.0]{GeV}$) in the central region
$|\eta|<2.5$. The value of $P(\Delta \eta)$ at $\Delta \eta=0$ is the
probability for an event to fulfill the trigger condition. Events were
generated with cluster hadronisation and
$p_{\perp,\,\text{cut}}=\unit[0.5]{GeV}$ for $\sqrt{s}=\unit[7]{TeV}$.}
\label{fig:mc_trigger}
\end{center}
\end{figure*}

The additional trigger conditions investigated were to demand at least one 
charged particle with $p_\perp>\unit[0.5\,(1.0)]{GeV}$ in the central region 
$|\eta|<2.5$ ('trigger 1\,(2)'). The dependence of the gap rate on these
trigger 
conditions is shown in Fig.~\ref{fig:mc_trigger}.  The probability is defined 
such that the probability at  $\Delta \eta=0$ equals the probability for
fulfilling the
trigger condition. This leads to a downwards shift of the entire distribution.
Otherwise, the trigger only affects the probabilities for very large gaps
($\Delta \eta>5$).

\begin{figure*}
\begin{center}
\includegraphics{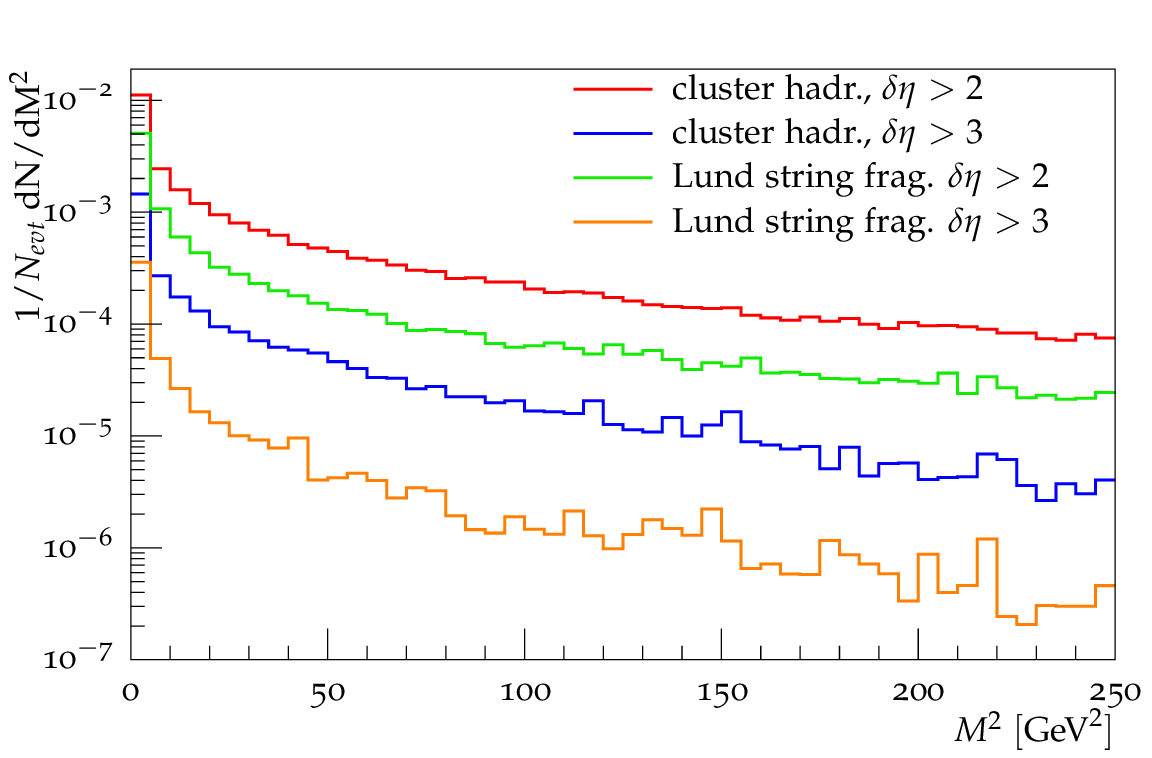}
\caption{\sf Distribution in squared invariant mass of the central system in
events with two rapidity gaps (definition 'all'), which are required to be both
larger than 2\,(3) units ($p_{\perp,\,\text{cut}}=\unit[0.5]{GeV}$, no trigger
condition,$\sqrt{s}=\unit[7]{TeV}$).}
\label{fig:mc_dpemass}
\end{center}
\end{figure*}

Given the large inclusive probabilities for rapidity gaps, fluctuations can also
fake a signature with two large gaps which usually is attributed to double
pomeron
exchange (DPE) processes. The probabilities for events with two gaps above 2 or
3 units are given in Table~\ref{tab:mc_dpeprobs}. Due to the requirement of two
large gaps these probabilities depend very strongly on the $p_\perp$ cut.
However, even relatively small rates can be dangerous since the DPE cross
section is much smaller than the inclusive QCD cross section. Furthermore, the
invariant mass of the system between the gaps has contributions out to large
masses (Fig.~\ref{fig:mc_dpemass}). The number of particles in the central
system also has a wide distribution that does not fall off very quickly.

\begin{table*}
 \begin{center}
  \begin{tabular}{|l|c|c|c|c|}
   \hline
   $p_{\perp,\,\text{cut}}$ & $\delta \eta>2$ cluster & $\delta \eta>2$ string
& $\delta \eta>3$ cluster & $\delta \eta>3$ string \\
   \hline
   $\unit[0.1]{GeV}$ & \unit[0.20]{\%} & \unit[0.03]{\%} & \unit[0.002]{\%} &
\unit[0.0001]{\%} \\
   $\unit[0.2]{GeV}$ & \unit[1.1]{\%} & \unit[0.12]{\%} & \unit[0.04]{\%} &
\unit[0.007]{\%} \\
   $\unit[0.5]{GeV}$ & \unit[17]{\%} & \unit[5.7]{\%} & \unit[1.5]{\%} & 
\unit[0.25]{\%} \\
   $\unit[1.0]{GeV}$ & \unit[55]{\%} & \unit[56]{\%} & \unit[19]{\%} & 
\unit[19]{\%}\\
   \hline
  \end{tabular}
  \caption{Probabilities for having two large gaps for the two different
hadronisation models at $\sqrt{s}=\unit[7]{TeV}$. All particles (charged and
neutral) in $|\eta|<5$ were considered. No trigger condition was required, but
there is no strong dependence on the trigger condition, since with two
large gaps the central system is likely to be in the tracking region
$|\eta|<2.5$.} \label{tab:mc_dpeprobs}
 \end{center}
\end{table*}

\subsection{Implication of fluctuations}

So, to study pure Pomeron exchange we {\it either} have to consider much larger 
gaps, where we are sure that the Pomeron dominates, {\it or}, in order to
extract 
the pure diffractive Pomeron-loop contribution, to study the $\delta \eta$
dependence 
of the cross section (of events with LRG) and to fit the data so as to subtract 
the part caused by the secondary Reggeons and/or by the fluctuations in the 
process of hadronization.

In principle this latter option sounds quite natural.  However, first note, the 
probability to observe LRG due to fluctuations in hadronization is rather large;
much larger than that expected from a simple Poissonian distribution. Moreover,
in some models the $\delta\eta$ dependence deviates from the simple exponential
form predicted by Poissonian behaviour, so 
it is  prohibitively hard to guarantee that Pomeron exchange is observed, 
rather than the details of models and their differences.  Indeed, 
given the crude analytical estimate and its order-of-magnitude confirmation by
the
MC study, it appears entirely possible that LRG of size $\delta \eta\simeq 3$ 
predominantly are caused by fluctuations. 

Therefore, first, we have to select larger gaps with $\delta \eta\sim 5$ or
more. 
Moreover, even fitting the gap size distribution by two (or more) exponents in
some 
model for hadronization, we may obtain an exponent with a large correlation
length 
see the lowest curve Fig.~\ref{fig:mc_had}. If the diffractive contribution is
about 5-10 mb this is not a problem. 
The long range-part coming from fluctuations is of the order of hundreds of 
$\mu$b (more than ten times less), but if the diffractive cross sections are
smaller 
than 1 mb then they will be extremely difficult to isolate.  In particular, the 
expected diffractive DPE cross sections are about 1 - 10 $\mu$b, while from 
fluctuations (assuming, for example, the ATLAS rapidity cuts \cite{ATLAS}) we
may obtain
up to 0.5 mb for the production of a central system of mass squared $M^2\sim 10$
GeV$^2$, with gaps $\delta\eta >2$ either side, see Fig. \ref{fig:mc_dpemass}. 
This immediately implies that 
there is practically no chance to unambiguously select {\it soft diffractive} 
DPE events based on LRG triggered in the calorimeter interval $|\eta|<5$. 
For this we would therefore need a larger $\eta$ coverage, which could be
achieved, 
for instance, by adding Forward Shower Counters \cite{fsc}.

Bearing in mind all theses problems it seems natural to prefer the first
definition
of diffraction, based on elastic scattering of the diffractive eigenstates. 
Unfortunately there is no way to directly measure low-mass diffractive
dissociation in 
the first years of LHC running.  Note that for low-mass dissociation at high
energies 
we deal with a very large rapidity gap $\delta\eta\gg 5$ and only the long
range 
correlation, caused by the Pomeron exchange, survives.  In this limiting case
the 
two definitions of diffraction become equivalent to each other\footnote{
  Here, let us recall, just for completeness, the expression for single-proton 
  dissociation, $p\to X$ corresponding to the triple-Reggeon diagrams, that is 
  the PPP (and PPR) diagram shown as the second term on the right hand side of 
  Fig. \ref{fig:epip}(c). When the mass $M_X$ of the system $X$ is small in 
  comparison with the initial energy $\sqrt{s}$, the dominant contribution
comes 
  from the Pomeron exchange, and the $M_X$ behaviour of the cross section takes 
  the form
  \begin{equation}
    \frac{\d\sigma^{\rm SD}}{\d M^2_X}\propto (M^2_X)^{\alpha_k-2\alpha_P}.
  \end{equation}
  The triple-Pomeron contribution (PPP term) has $\alpha_k=\alpha_P$, which, 
  for $\alpha_P  =1$, leads to $\d\sigma^{\rm SD}/\d M^2_X\sim 1/M^2_X$, whereas
for 
  the PPR term, which may be important at smaller mass, yields
  $\d\sigma^{\rm SD}/\d M^2_X\sim 1/M^3_X$.  Before the advent of the LHC,
triple-Reggeon 
  events were selected mainly by detecting forward protons with a large initial 
  momentum fraction $x_L$ close to 1. From a theoretical viewpoint this is the
same 
  as the selection of LRG; the size of the gap $\delta \eta\simeq
\ln(1/(1-x_L))$, and 
  the conventional choice $x_L>0.95$, are equivalent to $\delta \eta>3$.  At
relatively 
  low collider energies, from knowledge of the value of $x_L$, it was even
possible 
  to determine the mass of the diffractively produced system, $M^2_X=(1-x_L)s$, 
  and hence to distinguish the contributions of the different $N^*$ resonances. 
  At LHC energies, we have no hope of reaching such a good accuracy in the 
  measurement of $x_L$, and so we cannot study low-mass dissociation in this
way. 
  The only chance is to complement the LHC detectors by Forward Shower 
  Counters (FSC) \cite{fsc} in order to veto the production of extra secondaries
in 
  the region close to the fragmentation of the incoming proton, and thus to 
  suppress higher-mass dissociation.}.

\section{LRG in the early LHC runs}

Of course, the correlation length $l_f$ of fluctuations at the LHC energies may
be 
smaller than that measured at the Tevatron. Then the value of 
$\sigma_{\rm fluc}(\delta \eta>\Delta\eta)$ would not be so large. This
question should 
be studied experimentally in the first data runs of the LHC. Recall that the
early 
LHC runs, which have relatively low luminosity, are well suited for diffractive 
processes where the expected cross sections are rather large, and where pile-up
effects to not reduce the significance. Thus, we may select LRG events simply
by 
using the `veto' trigger, that is, by selecting events where in some interval
$\delta \eta > \Delta\eta$ no particles are observed\footnote{To suppress
fluctuations due to hadronization in non-diffractive events it would be better
to have a very low $p_{\perp\,\text{cut}}$, however then the fluctuations
in the calorimeter become important. In addition, it is not clear, a priori, how
hadronization effects impact on diffractive events.} with
$\pt>p_{\perp\,\text{cut}}$ in either the calorimeter or tracker\footnote{Note,
however, that even in high luminosity runs, we may study diffractive processes 
by selecting LRG events using the `veto' trigger. Of course, the efficiency of
the 
`veto' trigger becomes low at high luminosity, since quite often the gap will
be 
filled by the secondaries produced in the `pile-up' events.  If we denote the
mean 
number of inelastic $pp$ interactions per bunch crossing by $n$, then we have
the 
Poisson probability $P_n(0)=e^{-n}$ to have no additional `pile-up'
secondaries. 
Therefore the `veto' trigger actually acts at an effective luminosity, 
$L_{\rm eff}=L_0 e^{-n}$, which is much smaller than the true LHC luminosity
$L_0$. 
On the other hand, the expected diffractive cross sections are rather large.
For 
example, about $ 5\ -\ 10$ mb for single diffractive dissociation and a 
few $\mu$b for Double-Pomeron-Exchange (DPE) events with two LRG. Thus, even 
in the case of a pile-up of $n\sim 10$ the cross sections are sufficiently
large 
for the reduced luminosity, $L_{\rm eff}$, to reveal diffractive processes.}.

\subsection{Rapidity correlations}

Let us work in terms of the second definition, where the word `diffraction'
means 
Pomeron exchange. We will show how the multi-Pomeron effect can be studied at
the 
LHC, not only by selecting events with LRG, but also by measuring long-range
rapidity 
correlations. 

\subsubsection{Rapidity correlations in multiplicity, $R_2(N)$}

\begin{figure*} 
\begin{center}
\includegraphics[height=5cm]{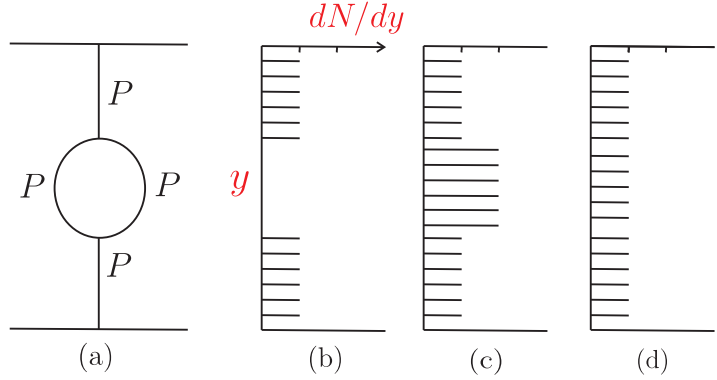}
\caption{\sf 
  Diagrams (b,c,d) show the particle density $\d N/\d y$ as a function of 
  rapidity $y$ for the four processes described by the Pomeron loop diagram
(a), 
  corresponding respectively, to neither, both or one Pomeron cut in the loop. 
  Diagram (d) occurs twice and corresponds to an absorption correction. The AGK 
  cutting rules give relative weights $1:2:-4$ to diagrams (b):(c):(d). }
\label{fig:4}
\end{center}
\end{figure*}

First note, from the AGK cutting rules \cite{agk}, that multi-Pomeron diagrams 
describe not only the processes with LRG, but simultaneously also events with a 
larger density of secondaries (when a few Pomerons are cut). The simplest
example 
is the four different cuts of the Pomeron loop diagram of Fig.
\ref{fig:loops}(b); 
the processes described by this diagram are shown in Fig. \ref{fig:4}.
Therefore, in 
two-particle inclusive cross sections, we should observe the same long-range
rapidity 
correlations, that is the same correlation length $l_P$, as in the LRG events. 

Let us explain this in a bit more detail:
If we were to cut $n$ Pomerons in a multi-Pomeron diagram, then we would get an 
event with multiplicity $n$ times larger than that generated by cutting just
one 
Pomeron.  The observation of a particle at rapidity $y_a$, say, has the effect
of 
enlarging the weight of the contribution of diagrams with many Pomerons cut.
For 
this reason the probability to observe another particle at quite a different 
rapidity $y_b$ becomes larger as well. This two-particle correlation can be
observed 
experimentally via the ratio of inclusive cross sections
\begin{equation}
R_2~=~
\frac{\sigma_{\rm inel}\d^2\sigma/\d y_a\d y_b}
     {(\d\sigma/\d y_a)(\d\sigma/\d y_b)}-1~=~
\frac{\d^2N/\d y_a\d y_b}{(\d N/\d y_a)(\d N/\d y_b)}-1\,,
\label{eq:R2}
\end{equation}
where $\d N/\d y=(1/\sigma_{\rm inel})~\d\sigma/\d y$ is the particle density.
Without 
multi-Pomeron effects the value of $R_2$ exceeds zero only when the two
particles are 
close to each other, that is when the separation $|y_a-y_b|\sim 1$ is not
large. 
Such {\it short-range} correlations arise from resonance or jet production. 

However, a positive value of $R_2$ at large $|y_a-y_b|$ will indicate the
presence 
of a {\it long-range} correlation arising from Pomeron loops.  To see this,
suppose 
that in a rapidity interval, which includes $y_a$ and $y_b$, there are $n$ cut 
Pomerons. That is, $n$ independent Multiple Interactions (MI) take place 
simultaneously in this interval. The total inelastic cross section 
\begin{equation}
\sigma_{\rm inel}=\sum_n \sigma_n\,,
\label{eq:n-mi1}
\end{equation}
the one-particle inclusive cross section
 \begin{equation}
\frac{\d\sigma}{\d y}=\sum_n n\cdot \sigma_n\cdot a\,,
\label{eq:n-mi2}
\end{equation}
and the two-particle inclusive cross section
\begin{equation}
\frac{\d^2\sigma}{\d y_a\d y_b}=\sum_n n^2\cdot \sigma_n\cdot a^2\,,
\label{eq:n-mi3}
\end{equation}
where $a=\d N^{(1)}/\d y$ is the particle density in an individual MI or cut
Pomeron.
Then the two-particle correlation, $R_2$ of (\ref{eq:R2}), takes the form 
\be
R_2=\frac{\langle n^2 \rangle}{\langle n \rangle ^2}-\ 1.
\ee 
Since $\langle n^2\rangle  > \langle n \rangle^2$ for the exchange of a few
Pomerons 
(i.e. a few MI), we expect $R_2>0$. Moreover, the rapidity interval occupied by
these 
Pomerons controls the correlation length measured via $R_2$. To be explicit,
the 
$R_2$ correlations measure the contributions due to those Pomeron loops which
embrace 
both $y_a$ and $y_b$.

\subsubsection{Rapidity correlations in $\et$, $R_2(\et)$}

Besides the correlation $R_2(N)$ between the particle densities (that is, the 
multiplicities) in different rapidity bins, we may measure the correlations
between 
the transverse momenta, $\pt$, of the secondaries (or correlations $R_2(\et)$,
of 
the transverse energy flow, $\d\et/\d\eta$).

If a non-zero correlation $R_2$ arises from Multiple Interactions and if
different 
MI do not depend on each other, then each Pomeron/MI (each parton shower)
should 
have the same $\pt$ distribution. Hence the correlations in $\et$ will be
identical 
to the multiplicity correlations, since in this case
\be
\frac{\d\et}{\d\eta}~=~\sum_i\sqrt{m_i^2+p^2_{i\,\perp}}~ \frac{ \d
N_i}{\d\eta},
\ee
and the $\pt$ distribution does not depend on the number of simultaneous 
Multiple Interactions. 

On the other hand, a correlation can arise from resonance decay or from jet 
production (and fragmentation), in an individual MI. These correlations are of 
short-range. They will result in a relatively narrow peak in the 
$|y_a-y_b|\sim {\cal O}(1)$ distribution. However, two high-$\et$ jets, which 
balance each other in $\pt$, may be separated by a relatively large rapidity 
interval. The correlation due to the production of such a high-$\et$  dijet
system 
is revealed better in the $\et$ flow. Moreover, a larger multiplicity at 
$\eta=y_a=y_{\rm{jet}1}$ should lead to a larger transverse energy flow, and a 
larger mean $\langle \pt\rangle$, in the region occupied by the other jet,
that 
is in the `away' azimuthal region (with respect to the first jet) with 
$\eta=y_b=y_{\rm{jet}2}$. We can always suppress these correlations, which 
originate from high-$\et$ dijets, by studying correlations in the `transverse' 
region in the azimuthal plane. The notation `away' and `transverse' is that of 
Field \cite{field}.

It would be most interesting to observe how the different MI {\it depend} on
each 
other. Indeed, we would not expect to have {\it independent} MI, each of which
are 
producing strongly interacting secondaries in a limited volume of configuration 
space. In an event with many MI, where the density of secondaries is large, it
is 
natural to expect a larger $\pt$ (due to the rescattering of the secondaries and
a stronger absorption of low $\pt$ hadrons; 
in other words, due to a larger saturation momentum\footnote{In Reggeon Field
Theory these effects 
are described by enhanced multi-Pomeron diagrams.}, $Q_s$). This effect will
result 
in a larger value of $\langle \pt\rangle$, or of the $\et$ flow, measured at 
$\eta=y_b$, in events with a larger particle density at  $\eta=y_a$. In terms
of 
the $\et$ flow, a correlation larger than that given by $R_2(N)$ would reflect 
the growth of $Q_s$ or/and $\langle \pt\rangle$  as the particle density,
increases. 
In other words we expect that $R_2(\et)>R_2(N)$.
 
At the LHC such long-range correlations can be measured in the ATLAS or CMS 
calorimeters which cover quite a large rapidity interval, $-5<\eta<5$.
The observation of similar rapidity distributions in both LRG events and $R_2$
correlations would be an argument in favour of the effects being due to Pomeron
exchange (that is, diffraction).

\subsection{Events with multiple rapidity gaps}
At the LHC, it would be informative to study not only the events with one LRG, 
but also events with two (or a few) LRG \cite{KMRprosp}. However, this would
require a significantly larger coverage in rapidity space. The gap survival 
probability $S^2_{\rm eik}$ corresponding to  {\it eikonal} rescatterings
(i.e.\ 
the probability that the gaps will not be filled by secondaries produced in an 
additional inelastic interaction of the incoming fast protons) depends very
weakly 
on the number of LRG in an event. On the other hand, the main absorptive effect 
may be due to {\it enhanced} rescattering involving intermediate partons. But, 
the number of different enhanced diagrams increases with the number of gaps,
and 
so, in this case, we may expect a stronger suppression of multi-gap events.
The simplest example is the Central Diffractive Production of some system $X$
of 
mass $M_X$ separated on both sides from the other secondaries by LRG, or even 
the `exclusive' process $pp\to p\oplus X\oplus p$ (where the $\oplus$ signs 
indicate LRG).  Such reactions are usually called Double-Pomeron-Exchange (DPE) 
processes.  It was demonstrated in \cite{nsoft} that the predictions for DPE 
cross sections depend strongly on the details of the model. Therefore the
detailed 
study of multi-gap events with much larger gaps at the LHC would select the most
realistic model of 
`soft' physics, and, moreover, constrain the values of the parameters used in
the 
model.

A more detailed discussion of soft physics at the LHC, and the qualitative
features 
that are essential for a realistic model of `soft' interactions, was presented
in 
\cite{softLHC}.

\section{Conclusions}

If we use the convention that {\it diffractive} events are just the `elastic'
scattering of `Good-Walker' eigenstates, then we consider only elastic
scattering and low-mass diffractive dissociation of the proton, processes which
are not immediately observable at the LHC.

From a wider viewpoint, any process due to Pomeron exchange may be called {\it
diffractive}. In general, such processes lead to Large Rapidity Gaps (LRG) in
the distribution of secondary hadrons. However, the probability to obtain a gap
without Pomeron exchange is not negligible; the gap can simply arise from
fluctuations in the hadronization process. The Monte Carlo studies presented in
this paper show that, with the present rapidity acceptances and $\pt$ cuts of
the LHC detectors, up to $\sim 0.5$ mb of the diffractive cross section can be
mimicked by fluctuations\footnote{As seen in fig.~\ref{fig:mc_had},
with $p_{\perp\, \text{cut}}=\unit[0.5]{GeV}$ the 
probability to have a gap $\delta\eta>5$ is between 0.01 (string hadronisation)
and 0.1 (cluster hadronisation). With the inelastic cross section
$\sigma_\text{inel}\sim \unit[50]{mb}$ this leads to $\unit[0.5 - 5]{mb}$ of LRG
caused by fluctuations.} which have nothing to do with Pomeron
exchange. This is
not a serious background if the cross section of the diffractive process that we
are studying is $\sim 10$ mb or larger, but it will pose a problem for studying
the so-called double-Pomeron exchange (DPE) events, where the expected cross
sections are $\sim 10~\mu$b.

In Ref.~\cite{CMSFWD-10-001} CMS\footnote{We thank the referee 
for pointing us to this note, which actually appeared after the submission of 
our manuscript.}
 claim observation of
inclusive diffraction at $\sqrt{s}=$ 900 and 2360 GeV.
The events have low multiplicity
 in the detector and/or have relatively low light-cone momentum, $E\pm p_z< 8$
GeV. In order to describe these events CMS use the single-diffractive
dissociation option in the PYTHIA or Phojet Monte Carlos.  The contribution
from the non-diffractive component (due to fluctuations) is about
3 - 4 times lower than that observed in the lowest multiplicity or the lowest
light-cone momentum bins. However, note that PYTHIA uses `string' hadronization
whereas for `cluster' hadronization the probability of fluctuations with $\Delta
\eta\sim 3$ is more than 3 times 
larger both with $p_{\perp,{\rm cut}}=0.1$ GeV and with $p_{\perp,{\rm
cut}}=0.5$ GeV, see Fig. \ref{fig:mc_had}. Therefore it is possible that the
main contribution, to the `diffractive' events 
observed by CMS,  may be actually due to fluctuations.
It would be interesting to study experimentally the dependence of the cross
section on the size of the rapidity gap, $\Delta\eta$, and on the value of
$p_{\perp,{\rm cut}}$, in order to constrain the model used for hadronization,
and hence to be able to select true diffractive events.

Recall, for processes driven by multi-Pomeron exchange, that the same diagram
describes LRG events and events with an enlarged multiplicity of secondaries.
Therefore, comparing the long-range correlations between multiplicities (and
between tranverse energy) with that observed in LRG events, it is possible to
confirm that the effect is due to Pomeron-exchange (and not due to
fluctuations).

\section*{Acknowledgements}
We thank Risto Orava and Andrew Pilkington for valuable discussions. MGR would
like 
to thank the IPPP at the University of Durham for hospitality. VAK thanks the
participants of the Diffraction at LHC meeting at CERN
for encouraging discussions. This work was 
supported by the Federal Programme of the Russian 
State RSGSS-3628.2008.2.


\end{document}